# Coagulapathies after vaccination against SARS-CoV-2 may be derived from a combination effect of SARS-CoV-2 spike protein and adenovirus vector-triggered signaling pathways

Ralf Kircheis[1]

[1] Syntacoll GmbH, Donaustrasse 24, 93342 Saal a.d. Donau, Germany

## Abstract

The novel coronavirus SARS-CoV-2 has resulted in a global pandemic with worldwide 6-digital infection rates and thousands death tolls daily. Enormeous effords are undertaken to achieve high coverage of immunization in order to reach herd immunity to stop spreading of SARS-CoV-2 infection. Several SARS-CoV-2 vaccines, based either on mRNA, viral vectors, or inactivated SARS-CoV-2 virus have been approved and are being applied worldwide. However, recently increased numbers of normally very rare types of thromboses associated with thrombocytopenia have been reported in particular in the context of the adenoviral vector vaccine ChAdOx1 nCoV-19 from Astra Zeneca. While statistical prevalence of these side effects seem to correlate with this particular vaccine type, i.e. adenonoviral vector based vaccines, the exact molecular mechanisms are still not clear. The present review summarizes current data and hypotheses for molecular and cellular mechanisms into one integrated hypothesis indicating that coagulopathies, including thromboses, thrombocytopenia and other related side effects are correlated to an interplay of the two components in the vaccine, i.e. the spike antigen and the adenoviral vector, with the innate and immune system which under certain circumstances can imitate the picture of a limited COVID-19 pathological picture.

## Introduction

The novel coronavirus SARS-CoV-2 (severe acute respiratory syndrome coronavirus 2) first reported in Wuhan, China, at the end of 2019 has developed into the heaviest global pandemic since the Spanish flu from 1918-1920, with worldwide more than 212 million infected persons and more than 4,4 million deaths by August 23rd, 2021, and 6-digit infection rates daily.

A major breakthrough in managing the COVID-19 pandemic has been the development and administration of several vaccines against SARS-CoV2. Meanwhile the European Medicines Agency (EMA) has approved four vaccines on the basis of randomized, blinded, controlled trials: two messenger RNA-based vaccines, *i.e.* BNT162b2 (BioNTech/Pfizer) and mRNA-1273 (Moderna) that encode the spike protein antigen of SARS-CoV-2 encapsulated in lipid nanoparticles; and two adenoviral vector-based vaccines, *i.e.* the ChAdOx1 nCov-19 (AZD1222, AstraZeneca), a recombinant chimpanzee adenoviral vector encoding the spike protein of SARS-CoV-2; and Ad26.COV2.S (Johnson & Johnson/Janssen), a recombinant human adenovirus type 26 vector encoding SARS-CoV-2 spike protein. A fifth vaccine, *i.e.* Gam-COVID-19-Vac ("Sputnik V", Gamaleya National Centre of Epidemiology and Microbiology, Russia) is under evaluation by the EMA. Furthermore, an inactivated SARS-CoV-2 vaccine developed by Sinovac, China, is being applied in various countries worldwide.

The approved vaccines have been extensively tested in large clinical trials with several thousands of volunteers and show high reactivity and efficacy of protection against severe COVID-19 and generally show good safety profiles. A panel of typical side effects, including pain at the injection site, fever, chills, fatique, and muscle pain have been reported, but no significant number of severe side effects has been reported from clinical studies (1-5).

However, recently reports of various types of venous thrombosis, and in particular of normally very rare **cerebral venous sinus thrombosis (CVST)** in a timely correlation to vaccination against SARS-CoV-2 with ChAdOx1 nCoV-19 have raised safety concerns. A variety of vaccine-associated thrombotic events including cerebral venous thrombosis, splanchnic-vein thrombosis, pulmonary embolism, and other thromboses, as well as disseminated intravascular coagulation have been reported in a time frame between a few days up to three weeks days after ChAdOx1 nCoV-19 vaccination against SARS-CoV-2 (6-8).

Among the various vaccine-associated thrombotic events special attention has been focussed on the normally very rare cerebral venous sinus thrombosis (CVST) in combination with pronounced thrombocytopenia, with most of these patients showing high levels of antibodies to platelet factor 4-polyanion complexes; without previous exposure to heparin pointing to a rare vaccine-related variant of spontaneous heparin-induced thrombocytopenia refered to as vaccine-induced immune thrombotic thrombocytopenia (VITT) (6-8).

A recent retrospective survey which estimated the incidence of CVST and other cerebrovascular events in temporal relation to COVID-19 vaccination with BNT162b2, ChAdOx1 nCoV-19, and mRNA-1273 in Germany showed an at least 10-fold higher CVST incidence rate in patients who received a first ChAdOx1 nCoV-19 vaccine shot compared with the highest estimate of CVST incidence rate from empirical data. Furthermore, an almost 10 fold higher risk for CVST following vaccination with ChAdOx1 nCov-19 compared to mRNA-based vaccines (9) together with recent reports on individuals who developed CVST with a severe thrombopenia within two weeks after immunization with Ad26.COV2.S (10, 11) suggest that VITT-associated thrombotic events may be associated with adenovirus vector-based vaccines coding for the SARS-CoV-2 spike protein indicating that the underlying mechanism relies on - or at least includes - the adenoviral vectors used in this vaccine type.

The underlying mechanism of action of these thrombotic events after adenoviral vector-based SARS-CoV-2 vaccines is still unknown. The present review summarizes the published data related to thrombotic events and different hypotheses for the VITT associated thrombotic events, and presents an integrated model indicating that both, SARS-CoV-2 spike protein and adenovirus vector can trigger signaling pathways which individually - and at a higher probability in combination - may trigger thromboses and thrombocytopenia following vaccination.

## SARS-CoV-2 and COVID-19 – the virus and the disease

SARS-CoV-2 belongs to enveloped positive-sense, single-stranded RNA viruses, similar to the two other highly pathogenic coronaviruses, SARS-CoV and Middle East Respiratory Syndrome (MERS-CoV) (12). SARS-CoV-2 binds to the angiotensin-converting enzyme-related carboxypeptidase-2 (ACE-2) receptor on the target cells by its spike (S) protein. The spike (S) protein is composed of the S1 subunit containing the highly conserved receptor binding domain (RBD) and the S2 subunit, which mediates fusion between the viral and host cell membranes after cleavage by the cellular serine protease TMPRSS2 (13). This furin-like cleavage site is unique to the S protein of SARS-CoV-2 and may together with the particularly high binding affinity to the target receptor and the peculiarity of a long symptom-free but nevertheless highly infectious time period between infection and appearance of first symptoms or asymptomatic transmission be responsible for the particularly efficient spread of SARS-CoV-2 compared to previous pathogenic hCoVs (14). The ACE-2 receptor is widely expressed in pulmonary and cardiovascular tissues and hematopoietic cells, including monocytes and macrophages, which may explain the broad range of pulmonary and extrapulmonary effects of SARS-CoV-2 infection, including cardiac, gastrointestinal organs, and kidney affection (13-15).

The majority of individuals infected with SARS-CoV-2 show mild-to moderate symptoms, and up to 20% of infections may be asymptomatic. Symptomatic patients show a wide spectrum of clinical manifestations ranging from mild febrile illness and cough up to acute respiratory distress syndrome (ARDS), multiple organ failure, and death. Thus, the clinical picture of severe cases is very similar to that seen in SARS-CoV- and MERS-CoV-infected patients (12). While younger individuals show predominantly mild-to-moderate clinical symptoms, elderly individuals frequently exhibit severe clinical manifestations (17-23). Pre-existing comorbidities, including diabetes, respiratory and cardiovascular diseases, renal failure and sepsis, higher age and male sex seem to be associated with more severe disease and higher mortality (20-24). Postmortem analysis of fatal COVID-19 showed diffuse alveolar disease with capillary congestion, cell necrosis, interstitial edema, platelet-fibrin thrombi, and infiltrates of macrophages and lymphocytes (25). Furthermore, induction of endotheliitis in various organs (including lungs, heart, kidney and intestine) by SARS-CoV-2 infection as a direct consequence of viral involvement and of the host inflammatory response has been demonstrated (15, 16).

The molecular mechanisms for the morbidity and mortality of SARS-CoV-2 are still incompletely understood. Virus-induced cytopathic effects and viral evasion of the host immune response, in particular the inhibition of the host IFN type I response by the SARS-CoV-2 (26), seem to play a role in disease severity. Furthermore, clinical data from patients, in particular those with severe clinical manifestations, indicate that highly dysregulated exuberant inflammatory and immune responses correlate with the severity of disease and lethality (15, 16, 18, 27-29). Significantly elevated cytokine and chemokine levels, also termed "cytokine storm", are assumed to play a central role in severity and lethality in SARS-CoV-2 infections. Elevated plasma levels of IL-1b, IL-7, IL-8, IL-9, IL-10, G-CSF, GM-CSF, IFNg, IP-10, MCP-1, MIP-1a, MIP-1b, PDGF, TNFa, and VEGF have been reported in both ICU (intensive care unit) patients and non-ICU patients. Notably, significantly higher plasma levels of IL-2, IL-7, IL-10, G-CSF, IP-10, MCP-1, MIP-1a, and TNFa were found in patients with severe pneumonia developing ARDS and requiring ICU admission and oxygen therapy compared to non-ICU patients showing pneumonia without ADRS (18). Various studies have shown that highly stimulated epithelial-immune cell interactions lead to exuberant dysregulated inflammatory responses with significantly (topically and systemically) elevated cytokine and chemokine release (30, 31).

Regarding the underlying signaling pathways, recent data suggest that the NF-kB pathway is one of the central signaling pathways for the SARS-CoV-2 infection-induced proinflammatory cytokine/ chemokine response, playing a central role in the severity and lethality of COVID-19 (32-38). This NF-kB-triggered proinflammatory response in acute COVID-19 is shared with other acute respiratory viral infections caused by highly pathogenic influenza A virus of H1N1 (e.g. Spanish flu) and H5N1 (avian flu) origin, SARS-CoV and MERS-CoV (38).

Excessive activation of exuberant inflammatory responses with involvement of endothelial cells, epithelial cells and immune cells are assumed to lead to further disturbances in a variety of other integrated systems, including the complement system, coagulation, and bradikinine systems, leading to increased coagulopathies and feeding back into positive signaling feedback loops accelerating COVID-19-associated inflammatory processes (39-44). In particular, vascular occlusion by neutrophil extracellular traps (NET) and disturbances of coagulation with various types of thromboses and multiple micro thromboses seem to be another hallmark of COVID-19 disease and development of coagulopathies is one of the key and persistent features associated with poor outcome. In particular elevated D-dimer levels, prolonged prothrombin time, and thrombocytopenia, together with low fibrinogen (indicating fibrinogen consumption) have been found as prognostic indicators for poor outcome (45-49). Lung histopathology often reveals fibrin-based blockages in the small blood vessels of patients who succumb to COVID-19 (25).

Furthermore, various types of antiphospholipid (aPL) antibodies targeting phospholipids and phospholipid-binding proteins including anti-cardiolipin IgG, IgM, and IgA; anti-β2 glycoprotein I IgG, IgM, and IgA; and anti-phosphatidylserine/prothrombin (aPS/PT) IgG and IgM were found in 52% of serum samples from 172 patients hospitalized with COVID-19. Higher titers of aPL antibodies were associated with neutrophil hyperactivity, including the release of neutrophil extracellular traps (NETs), higher platelet counts, more severe respiratory disease, and lower clinical estimated glomerular filtration rate. Similar to IgG from patients with anti-phospholipid syndrome, IgG fractions isolated from patients with COVID-19 promoted NET release from neutrophils isolated from healthy individuals. Furthermore, injection of IgG purified from COVID-19 patient serum into mice accelerated venous thrombosis in two mouse models. These findings suggest that half of patients hospitalized with COVID-19 become at least transiently positive for aPL antibodies and that these auto-antibodies are potentially pathogenic (**50**). High rates of thrombosis and thrombotic-related complications have been reported in adult patients with severe COVID-19 as well as in children developing COVID-19 or multisystem inflammatory syndrome (MIS-C). Studies in adults have invoked thrombotic micro angiopathy (TMA) as a potential cause for severe manifestations of COVID-19 (**51-53**). TMA results from endothelial cell damage to small blood vessels, leading to hemolytic anemia, thrombocytopenia, and, in some cases, organ damage (**54-58**). TMA has been reported in postmortem studies of adult patients with COVID-19 (**59**). Regarding therapeutic intervention, a retrospective analysis examined the association of in-hospital anticoagulation procedures with mortality, intubation, and major bleeding. In-hospital anticoagulation was associated with lower mortality and intubation among hospitalized COVID-19 patients (**60**).

**Frequently decribed side effects after vaccination against SARS-CoV-2**

Before certification by the regulatory authorities, i.e. FDA and EMA respectively, the vaccines had been tested in thousands of volunteers in large clinical trials (**1-5, 61**). A panel of typical side effects have been reported, such as short-term, mild-to-moderate pain, redness, and swelling at the injection site and systemic flu-like symptoms including fatigue, headache, muscle pain, chills, joint pains, and fever shown at varying degrees for all vaccines. There were no major differences between the mRNA vaccine and the adenoviral vector vaccines, with one exception, *i.e.* that the mRNA vaccine (*i.e.* BNT162b2 and mRNA-1273) showed more pronounced side effects after the second immunization and whereas the adenoviral vector vaccines showed more pronounced side effects after the first immunization, with lower intensity/prevalence of side effects after the second vaccination for ChAdOx1 nCoV-19 vaccine (AZD1222). Furthermore, younger individuals (<55 hears) showed genereally a higher incidence and intensity of side effects compared to aged persons (>55 years) reported for both types of vaccines, mRNA and adenoviral vector-based.

Importantly, no significant increase in prevalence of thrombotic events has been reported during the clinical studies, with large Phase 3 trials tested in more 30.000 – 40.000 volunteers the various vaccines (**1-6, 61**).

## Rare serious side effects after vaccination visible during mass vaccination after market approval

Following mass vaccination after market approval recent reports of cerebral venous sinus thrombosis (CVST) and a variety of other thrombotic events after ChAdOx1 vaccination against SARS-CoV-2 have raised safety concerns.

**One study recently published in N Eng J Med** showed venous thrombosis and thrombocytopenia seven to 10 days after receiving the first dose of the ChAdOx1 nCoV-19 adenoviral vector vaccine against coronavirus disease 2019 (Covid-19). All patients had high levels of antibodies to platelet factor 4 (PF4)-polyanion complexes; without previous exposure to heparin pointing to a rare vaccine-related variant of spontaneous heparin-induced thrombocytopenia refered to as vaccine-induced immune thrombotic thrombocytopenia (6). **A second study published in N Eng J med** assessed the clinical and laboratory characteristics of patients who had developed thrombosis or thrombocytopenia after vaccination with ChAdOx1 nCov-19 or other vaccine-associated thrombotic events, including 9 cerebral venous thrombosis, 3 splanchnic-vein thrombosis, 3 pulmonary embolism, and 4 other thromboses; 5 disseminated intravascular coagulation. All 28 patients tested positive for antibodies against PF4-heparin and tested positive in a platelet-activation assay in the presence of PF4 independent of heparin. Platelet activation was inhibited by high levels of heparin, Fc receptor-blocking monoclonal antibody, and immune globulin. Additional studies with PF4 or PF4-heparin affinity purified antibodies in 2 patients confirmed PF4-dependent platelet activation. The authors concluded that vaccination with ChAdOx1 nCov-19 can result in the rare development of immune thrombotic thrombocytopenia mediated by platelet-activating antibodies against PF4, which clinically mimics autoimmune heparin-induced thrombocytopenia (7). **A third study** published in New Engl J Med showed similar results in 22 patients presented with thrombocytonpenia and thrombosis, primarily cerebral veneous thrombosis, and 1 patient with isolated thrombocytopenia and a hemorrhagic phenotype. All of the patients had low or normal fibrinogen levels and strongly increased D-dimer levels, and 22 patients were positive for antibodies against PF-4, and 1 patient was negative (8). Meanwhile in several European countries, cases of venous thrombosis including cerebral venous sinus thrombosis (CVST) have been reported in temporal context with ChAdOx1 nCov-19 vaccine administration. At the beginning of March 2021, 30 venous thrombo-embolic events were reported to EMA out of about 5 million persons who had received the ChAdOx1 nCoV-19 vaccine at that time (9). The UK's Medicines and Healthcare Products Regulatory Agency had received 79 reports of thrombosis associated with low platelets by 31 March, of which 44 were CVST. Of these 79 cases, 51 (13 fatal) were in women and 28 (six fatal) in men. All of the UK cases occurred after the first dose. The risk was higher in the younger age groups, starting at 1.1 serious harm events for 100 000 immunised people among those aged 20-29 years and falling to 0.2/100 000 in those aged 60-69. These events were recorded within a time interval of less than 4 weeks after vaccination. For comparison, in women taking hormonal contraceptives the risk of thrombosis is about 60/100 000 person years and risk of fatal pulmonary embolism is about 1/100 000. Furthermore, several cases of serious thrombosis with thrombocytopenia have been reported after the use of the Johnson & Johnson (Janssen) covid-19 vaccine (62).

A recent retrospective survey estimated the incidence of CVST and other cerebrovascular events in temporal relation to COVID-19 vaccination with BNT162b2, ChAdOx1 nCov-19, and mRNA-1273 in Germany. According to this study up to April 14, 2021 Germany identified 62 vascular cerebrovascular adverse events in close temporal relationship with a COVID-19 vaccination, of which 45 cases were CVST. Eleven patients died. The authors estimated an incidence rate of CVST within one month from first dose administration of 17.9 per 100,000 person-years for ChAdOx1 nCov-19 vaccine and 1.3 per 100,000 person-for BNT162b2. Before the COVID-19 pandemic, the incidence rate of CVST has been estimated between 0.22 – 1.75 per 100,000 person-years in four European countries, Australia, Iran and Hong Kong. Accordingly, a 10 to 90 fold higher CVST incidence rate in patients who received a first ChAdOx1 nCov-19 vaccine shot compared with the highest or lowest estimate of CVT incidence rate from empirical data, respectively. The incidence rate of a CVST event after first dose COVID-19 vaccination was also statistically significantly increased for ChAdOx1 nCov-19 compared to mRNA-based vaccines (9.68, 3.46 to 34.98) and for females compared to non-females(3.14, 1.22 to 10.65) (9). The 10 fold higher risk for CVST following vaccination with ChAdOx1 nCov-19 compared to mRNA-based vaccines together with recent reports on individuals who developed CVST with a severe thrombopenia within two weeks after immunization with Ad26.COV2.S (10, 11, 62) suggests that VITT-associated thrombotic events may be associated with adenovirus vector-based vaccines coding for the SARS-CoV-2 spike protein indicating that the mechanism of action relies or at least includes the adenoviral vector used in this vaccine. The underlying mechanism of action of these thrombotic events after adenoviral vector-based SARS-CoC-2 vaccines is still not completely known, although various data and hypotheses have been raised (see below).

Beside vaccine-associated thrombotic events several additional serious conditions have been reported in association with vaccination against SARS-CoV-2, including single cases of capillary leakage syndrome and of coronary myocarditis after immunization with the ChAdOx1 nCov-19 (AZD1222) (AstraZeneca) and BNT162b2 (BioNTech/Pfizer), respectively.

## Superantigen Hypothesis

In order to elucidate the underlying mechanisms of various rare side effects following anti SARS-CoV-2 vaccination, one particular disease pattern appearing in a timely context to SARS-CoV-2 pandemics may be of interest. During the COVID-19 pandemic a new deadly disease in children named multisystem inflammatory syndrome in children (MIS-C) has got much attention, which rapidly progresses to hyperinflammation, shock and can lead to multiple organ failure in a high percentage of affected children. MIS-C has been found temporally associated with COVID-19 pandemic with a few weeks delay following peaks in SARS-CoV-2 infection incidence and was found often associated with SARS-CoV-2 exposure or presence of SARS-CoV-2 reactive antibodies in affected children. After initial reports in the UK, an increasing number of cases has been reported in Europe and New York, a few weeks after epidemic peaks, respectively. MIS-C manifests as persistent high fever, hyperinflammation with multiorgan system involvement including cardiac, gastrointestinal, renal, hematologic, dermatologic, and neurologic symptoms. The overall clinical picture of MIS-C, however, is often similar in many aspects to the late, severe CODIV-19 phase in adults, characterized by cytokine storm, hyperinflammtion, and multiorgan damage, severe myocarditis and acute kidney injury (**63**). A causal link between SARS-CoV-2 infection and MIS-C has not yet been firmly established; however, many patients with MIS-C were reportedly exposed to someone known or suspected to have COVID-19. Furthermore, only around a third of patients with MIS-C were tested positive for SARS-CoV-2 by PCR, but a large majority of PCR-negative were positive serologically for SARS-CoV-2 antibodies and/or had a history of mild COVID-19 infection or exposure several weeks before presentation. Such timing suggests that MIS-C is a postinfectious disease or an immune or autoimmune disease triggered by SARS-CoV-2 infection. Although initially MIS-C was seen to resemble Kawasaki diseases (KD) clinical, and laboratory characteristics indicate that MIS-C is rather reminiscent to the **toxic shock syndrome (TSS) usually found in severe cases after sepsis with Gram-positive bacteria such as** *Staphylococcus aureus* or *Streptococcus pyogenes* as indicated by typical gastrointenstinal involvement, myocardial disfunction and cardiovascular shock, pronounced lymphopenia and thrombocytopenia, and high coagulation parameters, such as D-dimers – found in MIS-C and TSS, but typically not in KD (**63**). Notably, TSS is known to be caused by different types of superantigens (SAgs) including bacterial and viral, with the bacterial SAgs being broadly studied. They include proteins secreted by ***Staphylococcus aureus and Streptococcus pyogenes*** that stimulate massive production of inflammatory cytokines and toxic shock. Typical examples are toxic shock syndrome (TSS) toxin 1, and staphylococcal enterotoxins B (SEB) and H (SEH). They are highly potent T cell activators that can bind to major histocompatibility complex (MHC) class II (MHCII) molecules and/or directly to T cell receptors (TCRs) of both CD4+ and CD8+ T cells. The ability of SAgs to bypass the antigen specificity of the TCRs results in broad activation of T cells and a cytokine storm, leading to toxic shock. Notably, SAgs do not bind the major (antigenic) peptide-binding groove of MHCII, but instead bind other regions or the αβTCRs, directly (63).

Importantly, using structure-based computational models, **Cheng et al demonstrated** that the SARS-CoV-2 spike (S) glycoprotein exhibits a high-affinity motif for binding TCRs, and may form a ternary complex with MHCII. The binding epitope on the spike protein harbors a sequence motif unique to SARS-CoV-2 (which is not present in other SARS-related coronaviruses), which is highly similar in both sequence and structure to the bacterial superantigen staphylococcal enterotoxin B. Furthermore, the interfacial region includes selected residues from an intercellular adhesion molecule (ICAM)-like motif shared between the SARS viruses from the 2003 and 2019 pandemics. A neurotoxin-like sequence motif on the receptor-binding domain also exhibits a high tendency to bind TCRs. Analysis of the TCR repertoire in adult COVID-19 patients demonstrated that those with severe hyperinflammatory disease exhibit TCR skewing consistent with superantigen activation (64). A blood test that determines the presence of specific TCR variable gene segments for identification of patients at risk for severe MIS-C has been developed (65). These data suggest that SARS-CoV-2 Spike protein itself may act as a superantigen to trigger the development of MIS-C as well as cytokine storm in adult COVID-19 patients (66). Interestingly, first cases of MIS following SARS-CoV-2 infection in adulty have been reported (67).

## Superantigen induce procoagulant activity

In the context of the superantigen hypothesis it is noteworthy that during severe sepsis activation of blood coagulation plays a critical pathophysiological role resulting in septic shock, microthrombi and multi organ dysfunction. During severe sepsis and septic shock, a massive release of cytokines and activation of the coagulation system result in **disseminated intravascular coagulation (DIC)** and multiorgan dysfunction syndrome (68-70). The procoagulant activity and tissue factor induction by various superantigens from *Staphylococcus aureus*, including enterotoxin A (EA), enterotoxin B (EB), and toxic shock syndrome toxin (TSST)-1, were tested for their ability to induce procoagulant activity and tissue factor (**TF**) expression – a major initiator of blood coagulation expressend predominantly on monocytic cells and endothelial cells - in human whole blood and in peripheral blood mononuclear cells. Determination of clotting time showed that all, enterotoxin A, B and toxic shock syndrome toxin 1 from *S. aureus* induced procoagulant activity in whole blood and in mononuclear cells. The procoagulant activity was dependent on the expression of TF in monocytes. In the supernatants from staphylococcal toxin-stimulated mononuclear cells, interleukin (IL)-1 beta was detected by ELISA. The increased procoagulant activity and TF expression in monocytes induced by the staphylococcal toxins were inhibited in the presence of IL-1 receptor antagonist, a natural inhibitor of IL-1 beta. The study demonstrated that superantigens from *S. aureus* activate the extrinsic coagulation pathway by inducing expression of TF in monocytes, and that the expression is mainly triggered by superantigen-induced IL-1 beta release (**69, 70**).

### Superantigens & NF-kappaB

Furthermore, is has been demonstrated that **toxic shock syndrome toxin-1 (TSST-1) and staphylococcal enterotoxins A and B induce the activation of NF-κB**, that acts as a transcriptional enhancer by binding to sequences found in both the IL-1 beta and TNF-alpha promoters. Induction of both NF-κB DNA-binding proteins and NF-κB enhancer function was down-regulated by inhibitors of protein kinase C and protein tyrosine kinase, indicating a role for these protein kinases in the induction of NF-κB by MHC class II ligands. Using neutralizing antibodies, it was demonstrated that after the stimulation of cells with TSST-1, TNF-alpha acted to up-regulate binding of NF-κB to DNA and the activation of the NF-κB-promoter CAT construct indicating that induction of NF-κB by superantigens is up-regulated in part by an autocrine loop involving TNFα (**71**).

The central role of the NF-κB pathway in superantigen mediated T-cell activation has also been demonstrated in studies showing that proteasome inhibition reduced superantigen-mediated T cell activation. PS-519 as a potent and selective proteasome inhibitor was shown to inhibit NF-κB activation by blocking the degradation of its inhibitory protein IκB and to reduce superantigen-mediated T cell-activation *in vitro* and *in vivo*. Proliferation was inhibited along with the expression of very early (CD69), early (CD25), and late T cell (HLA-DR) activation molecules. Moreover, expression of E-selectin ligands relevant to dermal T cell homing was reduced, as was E-selectin binding *in vitro* (**72**). Furthermore, inhibition of NF-κB pathway by two anti-oxidants, N-acetyl-cysteine (NAC) and pyrrolidine dithiocarbamate (PDTC) was shown to dose-dependently inhibit SE-stimulated T-cell proliferation (by 98%), production of cytokines and chemokines by PBMC and expression of SE-induced cell surface activation markers. The potency of both NAC and PDTC corresponded to their ability to inhibit NF-κB activation (**73**). Beside the MHCII dependent activation of T-cells, also a MHC/II independent direct stimulation of TCR Vb by *Staphylococcus* aureus enterotoxin via PKCtheta/ NF-κB and IL2R/STAT signaling pathways has been shown (**74**).

## NF-κB pathway is part of normal T cell activation

In this context it has to be mensioned that normal or physiological antigen stimulation of TCR signaling to NF-κB is required for T cell proliferation and differentiation of effector cells. Engagement of the TCR by an MHC-antigen complex initiates a whole chain of downstream events, described in detail by Paul et al., which ultimately trigger calcium release and PKC activation, respectively. Activation of a specific PKC isoform, PKCθ, connects TCR proximal signaling events to distal events that ultimately lead to NF-κB activation. Importantly, PKCθ activation is also driven by engagement of the T cell costimulatory receptor CD28 by B7 ligands on antigen presenting cells and via intermediate steps leads to the activation of IKKβ. IKKβ then phosphorylates IκBα, triggering its proteasomal degradation, enabling nuclear translocation of canonical NF-κB heterodimers comprised of p65 (RELA) and p50 proteins. Once in the nucleus, NF-κB governs the transcription of numerous genes involved in T cell survival, proliferation, and effector functions (75).

## SARS-CoV-2 Spike protein induces NF-κB

Importantly, the pathogenesis of COVID-19 has been shown to involve over-activation of NF-κB pathway (37, 38). In several studies it has been studied which part(s) of the SARS-CoV-2 is responsible for the massive NF-κB pathway activation. Khan et al investigated direct inflammatory functions of major structural proteins of SARS-CoV-2 and showed that spike (S) protein potently induces inflammatory cytokines and chemokines including IL-6, IL-1ß, TNFα, CXCL1, CXCL2, and CCL2, but not IFNs in human and mouse macrophages. No such inflammatory response was observed in response to membrane (M), envelope (E), and nucleocapsid (N) proteins. When stimulated with extracellular S protein, human lung epithelial cells A549 also produced inflammatory cytokines and chemokines. Interestingly, epithelial cells expressing S protein intracellularly are non-inflammatory, but elicited an inflammatory response in macrophages when co-cultured. Biochemical studies revealed that S protein triggers inflammation via activation of the NF-κB pathway in a MyD88-dependent manner. Further, such an activation of the NF-κB pathway was abrogated in TLR2-deficient macrophages. Consistently, administration of S protein induced IL-6, TNFα, and IL-1 ß in wild-type, but not TLR2-deficient mice. In this study, both S1 and S2 were demonstrated to show high NF-κB activation, with S2 showing the higher potency on an equimolar basis (76). **In a second study the** spike protein was demonstrated to promote an angiotensin II type 1 receptor (AT1) mediated signaling cascade, induced the transcriptional regulatory molecules NF-κB and AP-1/c-Fos via MAPK activation, and increased IL-6 release (77). **A third study** has demonstrated that SARS-CoV-2 spike protein subunit 1 (CoV2-S1) induces high levels of NF-κB activations, production of pro-inflammatory cytokines and mild epithelial damage, in human bronchial epithelial cells. CoV2-S1-induced NF-κB activation requires S1 interaction with human ACE2 receptor and early activation of endoplasmic reticulum (ER) stress, and associated unfolded protein response (UPR), and MAP kinase signaling pathways. The FDA-approved ER stress inhibitor, 4-phenylburic acid (4-PBA), and MAP kinase inhibitors, trametinib and ulixertinib, ameliorated CoV2-S1-induced inflammation and epithelial damage (78). **In a fourth study**, the effects of a recombinant SARS-CoV-2 spike glycoprotein S1 was investigated in human peripheral blood mononuclear cells (PBMCs). Stimulation of PBMCs with spike glycoprotein S1 (100 ng/mL) resulted in significant increase in TNFα, IL-6, IL-1β and IL-8. Pre-treatment with dexamethasone (100 nM) caused significant reduction in the release of these cytokines. Further experiments revealed that S1 stimulation of PBMCs increased phosphorylation of NF-κB p65 and IκBα, and IκBα degradation. DNA binding of NF-κB p65 was also significantly increased following stimulation with spike glycoprotein S1. Treatment of PBMCs with dexamethasone (100 nM) or the specific NF-κB inhibitor BAY11-7082 (1 µM) resulted in inhibition of spike glycoprotein S1-induced NF-κB activation. Activation of p38 MAPK by S1 was blocked in the presence of dexamethasone and SKF 86002. CRID3, but not dexamethasone pre-treatment produced significant inhibition of S1-induced activation of NLRP3/caspase-1. Further experiments revealed that S1-induced increase in the production of TNFα, IL-6, IL-1β and IL-8 was reduced in the presence of BAY11-7082 and SKF 86002, while CRID3 pre-treatment resulted in the reduction of IL-1β production. These results suggest that SARS-CoV-2 spike glycoprotein S1 stimulated PBMCs to release pro-inflammatory cytokines through mechanisms involving activation of NF-κB, p38 MAPK and NLRP3 inflammasome (79).

Furthermore, an interaction between SARS-CoV-2 spike (S) protein and LPS was shown to lead to aggravated inflammation *in vitro* and *in vivo*. Native gel electrophoresis demonstrated that SARS-CoV-2 S protein binds to LPS. Microscale thermophoresis yielded a KD of ∼ 47 nM for the interaction. Computational modeling identified a main LPS-binding site in SARS-CoV-2 S protein. S protein, when combined with low levels of LPS, boosted NF-κB activation in monocytic THP-1 cells and cytokine responses in human blood and peripheral blood mononuclear cells, respectively. The *in vitro* inflammatory response was further validated by employing NF-κB reporter mice and *in vivo* bioimaging. Dynamic light scattering, transmission electron microscopy, and LPS-FITC analyses demonstrated that S protein modulated the aggregation state of LPS, providing a molecular explanation for the observed boosting effect providing a potential molecular link between excessive inflammation during infection with SARS-CoV-2 and comorbidities involving increased levels of bacterial endotoxins (80).

Overall, these data show that SARS-CoV-2 *S* spike protein, induces powerful NF-kappaB activation showing strong similarity to data recorded for SARS-CoV *S* protein. Similar to SARS-CoV-2 also the clinical picture of severe acute respiratory syndrome (SARS) is characterized by an over-exuberant immune response with lung lymphomononuclear cells infilteration and proliferation that may account for tissue damage more than the direct effect of viral replication. Purified recombinant *S* protein was studied for stimulating murine macrophages (RAW264.7) to produce proinflammatory cytokines (IL-6 and TNFα) and chemokine IL-8. The authors found that a direct induction of IL-6 and TNF-release in the supernatant in a dose-, time-dependent manner and highly spike protein-specific. IL-6 and TNF-production were dependent on NF-κB, which was activated through IκB degradation (82).

## Platelet Factor 4

Vaccine-induced thrombotic thrombocytopenia *(VITT)* as observed in rare cases following vaccination with vaccines from AstraZeneca or Johnson & Johnson resembled Heparin-induced thrombocytopenia (HIT). The hallmark of HIT are antibodies to the heparin/platelet factor 4 (PF4) complex that cause thrombocytopenia and thrombosis through platelet activation.

Platelet factor 4 (PF4, CXCL4) is a small chemokine protein, present as positively charged tetramers, released by activated platelets. Its major physiological function is to promote blood coagulation. This function is related to PF4's affinity for heparin and other glycosaminoglycans (GAGs), which are long linear polysaccharides consisting of repeating disaccharide units. By neutralizing the negatively charged heparan sulfate side chains of GAGs on the surface of platelets and endothelial cells, PF4 facilitates platelet aggregation to form a thrombus. However, PF4 has additional activities beyond simply promoting blood coagulation, and the regulation of PF4 is very complex. PF4 expression is elevated following trauma with the physiological role to prevent blood loss from injury. Notably, a significant amount of PF4 is also released by activated platelets in response to infection (82).

After exposure of patients to heparin, it binds to PF4 and promotes PF4 aggregation, so that they form ultra-large PF4/heparin complexes with antigenic properties. Some patients develop antibodies against PF4/heparin complexes that cause HIT. Anti-PF4/heparin antibodies were detected in 3.1–4.4% of healthy subjects, but occur in ~27–61% of patients after cardiac surgery and in 8–17% of medical and surgical patients treated with heparin. It is estimated that only 5–30% of patients with antibodies against heparin develop HIT (82). The pathogenesis of HIT involves circulating PF4/heparin/antibody complexes that bind to the FcγRIIA receptor on platelets and other Fc-receptor-bearing blood cells, such as monocytes and neutrophils. Activation of FcγRIIA causes platelet activation which leads to secretion of the contents of their cytoplasmic granules and to generation of procoagulant microparticles. In addition, platelet–neutrophil interactions triggered by HIT antibodies may activate vascular endothelium. PF4/heparin immune complexes also directly activate endothelial cells without involving FcγRIIA, inducing enhanced expression of adhesion molecules such as P- and E-selectins and the release of von Willebrand factor. The combination of direct platelet activation by HIT-related immune complexes through FcγRIIA and transactivation by monocyte and likely endothelial cell-derived thrombin increases expression of phosphatidylserine (PS) and binding of factor Xa to platelets. These consequences lead to generation of thrombin which increases the risk for thrombotic vessel occlusions, such as venous thromboembolism, myocardial infarction, or stroke (83). Binding of HIT IgG occurs only over a narrow molar ratio of reactants, being optimal at 1 mol PF4 tetramer to 1 mol unfractionated heparin (UFH). At these ratios, PF4 and UFH form ultralarge (>670 kD) complexes that bind multiple IgG molecules/complex, are highly antigenic, and promote platelet activation. Low molecular weight heparin (LMWH), which is less antigenic, forms ultralarge complexes less efficiently and largely at supra-therapeutic concentrations. In transgenic mice that vary in expression of human PF4 on their platelets, antigenic complexes form between PF4 and endogenous chondroitin sulfate. Binding of HIT IgG to platelets and induction of thrombocytopenia *in vivo* is proportional to PF4 expression. Heparin prolongs the duration and exacerbates the severity of the thrombocytopenia. In contrast, high doses of heparin, as used in CPB, or protamine, which competes with PF4 for heparin, disrupts antigen formation and prevents thrombocytopenia induced by HIT antibody (84).

A most striking observation, however, was that even heparin-naive patients can generate anti-PF4/heparin IgG as early as day 4 of heparin treatment, suggesting pre-immunization by antigens mimicking PF4/heparin complexes. These antibodies have been suggested to result from bacterial infections, as indicated by at least three lines of evidence: *(i)* PF4 bound charge-dependently to various bacteria, *(ii)* human heparin-induced anti-PF4/heparin antibodies cross-reacted with PF4-coated *Staphylococcus aureus* and *Escherichia coli*; and *(iii)* mice developed anti-PF4/heparin antibodies during polymicrobial sepsis without heparin application. Thus, after binding to bacteria, the endogenous protein PF4 induces antibodies with specificity for PF4/polyanion complexes. These can target a large variety of PF4-coated bacteria and enhance bacterial phagocytosis *in vitro*. The same antigenic epitopes are expressed when pharmacologic heparin binds to platelets augmenting formation of PF4 complexes. Boosting of preformed B cells by PF4/heparin complexes could explain the early occurrence of IgG antibodies in HIT. A continuous, rather than dichotomous, distribution of anti-PF4/heparin IgM and IgG serum concentrations in a cross-sectional population study (n = 4029) was found, indicating frequent pre-immunization to modified PF4. Thus, PF4 may have a role in bacterial defense, and HIT is probably a misdirected antibacterial host defense mechanism (**85**).

We hypotized that cationic PF4 may charge-dependently associate not only with bacterial surface structures, but also with viruses harboring a negative surface charge such as adenovirus. Due to the highly negatively charged hexone protein – which represents the major capsid protein - adenovirus expose a highly negative surface charge (**86, 87**). In this context, complexation with polycations has been used to enhance transfection efficacy of adenoviral vectors into target cells. It has been demonstrated that that a cationic component can charge-associate with adenovirus particles, which carry a net negative surface charge, and will facilitate attachment to the negatively charged cell membrane – an approach employed for increasing gene transfer efficacy of adenoviral vectors (**88**). Similiarly adsorption of Ad2 beta gal2 in the presence of different polycations, such as polybrene, protamine, DEAE-dextran, or poly-L-lysine significantly increased transfection efficacy into various cell types. The polyanion heparin completely abrogated the effects of polycations (**89**). Finally, a recent bioRxiv preprint has revealed the structure of the ChAdOx1/AZD-1222. ChAdOx1 shares the archetypal icosahedral, T=25, capsid common to adenoviruses. Binding of fiber knob to the coxsackie and adenovirus receptor (CAR) was shown to be the primary mechanism ChAdOx1 uses to attach to cells. Further, the work revealed that the surface of the ChAdOx1 viral capsid has a strong electronegative potential. The ChAdOx1 hexon hypervariable rgions (HVRs) are different to other adenoviruses, in terms of apical electrostatic surface potential, calculated on equilibrated hexon structures, ChAdOx1 has the most electronegative surface potential which can be expected to influence the strength of incidental charge-based interactions with other molecules. Molecular simulations have suggested that this charge, together with shape complementarity, are a mechanism by which an oppositely charged protein, e.g. platelet factor 4 (PF4) may bind the vector surface (**90**).

While a recent study by Greinacher's team has shown no cross-reactivity of platelet-activating antibodies isolated from VITT patients with the coronavirus spike protein (and only few out of 200 patients recovered from COVID-19 reacting weakly with PF4) (**91**) it would be – on the other side - interesting to test whether anti PF4 antibodies purified from patients developing CVST after vaccination with adenoviral-vector based vaccines show cross-reactivity with adenoviral vector used in the vaccines.

An alternative charge-based hypothesis has been proposed. Among the 50 billion or so virus particles in each dose, some may break apart and release their DNA. Like heparin, DNA is negatively charged, which could bind to the positively charged PF4. The complex might then trigger the production of antibodies, especially when the immune system is already on high alert because of the vaccine. An immune reaction to extracellular DNA is part of an ancient immune defense triggered by severe infection or injury, and free DNA itself can signal the body to increase blood coagulation (7, 92).

### Additonal activities of PF4

Activated platelets release micromolar concentrations of the chemokine CXCL4/Platelet Factor-4. Deposition of CXCL4 onto the vascular endothelium is involved in atherosclerosis, facilitating monocyte arrest and recruitment. CXCL4 was shown to drive chemotaxis of the monocytic cell line THP-1, and induced CCR1 endocytosis and monocyte chemotaxis in a CCR1 antagonist-sensitive manner (93). The role of the platelet chemokine platelet factor 4 (PF4) in hemostasis and thrombosis has been described (94).

### The impact of the adenoviral vector

One important factor contributing to the thrombotic events observed following vaccination with adenoviral vector-based anti SARS-CoV-2 vaccines may rely in the adenoviral vector used to deliver the transgene. The recombinant adenovirus ChAdOx1 used in the ChAdOx1 nCoV-19 vaccine against SARS-CoV-2 is replication-defective in normal cells. 28 kbp of adenovirus genes is delivered to the cell nucleus alongside the SARS-CoV-2 S glycoprotein gene. A recent study used direct RNA sequencing to analyse transcript expression from the ChAdOx1 nCoV-19 genome in human MRC-5 and A549 cell lines that are non-permissive for vector replication alongside the replication permissive cell line, HEK293. The expected SARS-CoV-2 S coding transcript dominated in all cell lines. However, they also detected rare S transcripts with aberrant splice patterns or polyadenylation site usage. Adenovirus vector transcripts were almost absent in MRC-5 cells, but in A549 cells, there was a broader repertoire of adenoviral gene expression at very low levels. Proteomically, in addition to S glycoprotein, multiple adenovirus proteins were detected in A549 cells compared to just one in MRC5 cells. The study demonstrated that low levels of viral backbone gene transcription takes place alongside very high levels of SARS-CoV-2 S glycoprotein gene transcription following SARS-CoV-2 vaccine ChAdOx1 nCoV-19 infection of human cell lines (95).

Adenoviral vectors are generally considered a safe and potently immunogenic vaccine delivery platform. Non-replicating Ad vectors possess several attributes which make them attractive vaccines for infectious disease, including their capacity for high titer growth, ease of manipulation, safety, and immunogenicity in clinical studies, as well as their compatibility with clinical manufacturing and thermo-stabilization procedures. In general, Ad vectors are highly immunogenic vaccines, which elicit robust transgene antigen-specific cellular (namely CD8$^+$ T cells) and/or humoral immune responses. It has been suggested that a combination of factors contribute to the potent immunogenicity of particular Ad vectors, including the magnitude and duration of vaccine antigen expression following immunization. Furthermore, the excessive induction of Type I interferons by some Ad vectors has been desscribed. Entry factor binding or receptor usage of distinct Ad vectors can affect their *in vivo* tropism following administration by different routes. The abundance and accessibility of innate immune cells and/or antigen-presenting cells at the site of injection contributes to early innate immune responses to Ad vaccination, affecting the outcome of the adaptive immune response (**96**).

However, intravenous (i.v.) delivery of recombinant adenovirus serotype 5 (Ad5) vectors for gene therapy is hindered by safety and efficacy problems. Most i.v.-delivered Ad5 is sequestered in the liver, and animal studies indicate that Kupffer cells (KCs) play a major role in this trapping. Studies have shown that liver sequestration is not mediated by the Ad5 receptor, CAR, but involves either a direct or a blood factor (coagulation factors IX and X and complement protein C4BP)-mediated interaction between the Ad fiber and cellular heparansulfate proteoglycans. Stone et al have discovered an additional pathway involved in unspecific Ad5 sequestration and degradation. After i.v. administration, Ad5 rapidly binds to circulating platelets, which causes their activation/aggregation and subsequent entrapment in liver sinusoids. Virus-platelet aggregates are taken up by Kupffer cells and degraded. Adenovirus sequestration in organs could be reduced by platelet depletion prior to vector injection (**97**).

Importantly, thrombocytopenia has been consistently reported following the administration of adenoviral gene transfer vectors. In one study, the authors have assessed the influence of von Willebrand Factor (VWF) and P-selectin on the clearance of platelets following adenovirus administration. In mice, thrombocytopenia occurs between 5 and 24 hours after adenovirus delivery. The virus activated platelets and induced platelet-leukocyte aggregate formation. There was an associated increase in platelet and leukocyte-derived microparticles. Adenovirus-induced endothelial cell activation was shown by VCAM-1 expression on virus-treated, cultured endothelial cells and by the release of ultra-large molecular weight multimers of VWF within 1 to 2 hours of virus administration with an accompanying elevation of endothelial microparticles. In contrast, VWF knockout (KO) mice did not show significant thrombocytopenia after adenovirus administration. They showed further that adenovirus interferes with adhesion of platelets to a fibronectin-coated surface and flow cytometry revealed the presence of the Coxsackie adenovirus receptor on the platelet surface. They conclude that VWF and P-selectin are critically involved in a complex platelet-leukocyte-endothelial interplay, resulting in platelet activation and accelerated platelet clearance following adenovirus administration (**98**).

Furthermore, replication-deficient adenoviruses are known to induce acute injury and inflammation of infected tissues, thus limiting their use for human gene therapy. The chemokine expression was evaluated in DBA/2 mice following the intravenous administration of various adenoviral vectors. Administration of adCMVbeta gal, adCMV-GFP, or FG140 intravenously rapidly induced a consistent pattern of C-X-C and C-C chemokine expression in mouse liver in a dose-dependent fashion. One hour following infection with 10(10) PFU of adCMVbeta gal, hepatic levels of MIP-2 mRNA were increased >60-fold over baseline. MCP-1 and IP-10 mRNA levels were also increased immediately following infection with various adenoviral vectors, peaking at 6 hr with >25- and >100-fold expression, respectively. Early induction of RANTES and MIP-1beta mRNA by adenoviral vectors also occurred, but to a lesser degree. The induction of chemokines occurred independently of viral gene expression since psoralen-inactivated adenoviral particles produced an identical pattern of chemokine gene transcription within the first 16 hr of administration. The expression of chemokines correlated as expected with the influx of neutrophils and CD11b+ cells into the livers of infected animals. At high titers, all adenoviral vectors caused significant hepatic necrosis and apoptosis following systemic administration to DBA/2 mice. To investigate the role of neutrophils in this adenovirus-induced hepatic injury, animals were pretreated with neutralizing anti-MIP-2 antibodies or depleted of neutrophils. MIP-2 antagonism and neutrophil depletion both resulted in reduced serum ALT/AST levels and attenuation of the adenovirus-induced hepatic injury histologically, confirming that this early injury is largely due to chemokine production and neutrophil recruitment (**99**).

Also gene transfer to the respiratory tract by replication-deficient adenoviruses is limited by the induction of inflammatory and immune responses. E1-E3-deleted recombinant adenovirus carrying the expression cassette for the cystic fibrosis gene (Ad.CFTR) were shown to upregulate the expression of the pro-inflammatory intercellular adhesion molecule-1 (ICAM-1) both *in vitro* and *in vivo*, and a central role for the NF-κB in Ad.CFTR-dependent up-regulation of ICAM-1 in respiratory epithelial A549 cells has been suggested. Specifically, Ad.CFTR induced translocation of NF-κB into the nucleus and binding to the proximal −228/−218 NF-κB consensus sequence on the ICAM-1 promoter. Ad.CFTR also stimulated a 13-fold increase in NF-κB-dependent expression of the CAT reporter gene under the control of a region of the ICAM-1 promoter, including the proximal NF-κB consensus sequence. The Ad.CFTR-dependent increase of ICAM-1 mRNA was abolished by inhibitors of NF-κB, such as N-acetyl-L-cysteine, pyrrolidine dithiocarbamate, parthenolide and the synthetic peptide SN50. All these inhibitors abolished both Ad.CFTR-induced NF-κB DNA binding and transactivating activities. These results have indicated a critical role of NF-κB in the pro-inflammatory response elicited by replication-deficient adenoviral vectors in respiratory cells (**100**). Furthermore, it was investigated whether the early phase of virus-cell interaction is sufficient to stimulate ICAM-1 upregulation. A549 cells were exposed to wild-type Ad5 (Ad5), to Ad.CFTR, and to Ad5 inactivated by incubation at 56 degrees C (Ad5/56 degrees C). All, Ad5, Ad.CFTR, and Ad5/56 degrees C activated NF-kappaB and increased ICAM-1 mRNA levels within 4 h after exposure. The role of the mitogen-activated protein kinases (MAPKs) on the ICAM-1 mRNA induction was studied. ICAM-1 mRNA upregulation was inhibited upon incubation with several chemicals, auch as ERK1/2 inhibitors PD98059 and AG1288 (by 98 and 67%, respectively), of the p38/MAPK pathway SB203580 (by 50%), of the JNK pathway dimethylaminopurine (by 83%), and of the NF-kappaB parthenolide (by 96%). Ad5 and Ad5/56 degrees C stimulated ERK1/2, p38/MAPK, and JNK1 starting 10 min and peaking 20-30 min after exposure. The present results indicate a link between the activation of the three major MAPK pathways, NF-kappaB, and the upregulation of ICAM-1 gene expression evoked by Ad5 in the very initial phase of infection (101). Adenovirus serotypes 2 or 5 (Ad2/5) enter respiratory epithelia after initial binding of fiber with the coxsackie-adenovirus receptor (CAR) or, alternatively, with cell surface heparan sulfate glycosaminoglycans. Ad2/5 internalization is triggered by binding of penton base to cellular RGD-binding integrins. The authors investigated the role of the Ad5 surface domain proteins constituting the vector capsid, *i.e.* the fiber, the penton base, and the hexon, on the transmembrane signals leading to the transcription of the different proinflammatory genes in the human respiratory A549 cell line. Interaction of Ad fiber with CAR activates both ERK1/2 and JNK MAPK and the nuclear translocation of NF-kappaB, whereas no activation was observed after exposing A549 cells to penton base and hexon proteins. Moreover, interaction of Ad fiber with CAR, but not heparan sulfate proteoglycans, promoted transcription of the chemokines IL-8, GRO-alpha, GRO-gamma, RANTES, and interferon-inducible protein 10. These results identified the binding of Ad5 fiber with the cellular CAR as a key proinflammatory activation event in epithelial respiratory cells that is independent of the transcription of Ad5 genes (102).

In another study, the vaccinal adjuvant role of rAd independently of its vector function was evaluated. BALB/c mice received one subcutaneous injection of a mixture of six lipopeptides (LP6) used as a model immunogen, along with AdE1 degrees (10(9) particles), a first-generation rAd empty vector. Although coinjected with a suboptimal dose of lipopeptides, AdE1 degrees significantly improved the effectiveness of the vaccination, even in the absence of booster immunization. In contrast to mice that received LP6 alone or LP6 plus a mock adjuvant, mice injected with AdE1 degrees plus LP6 developed both a polyspecific T-helper type 1 response and an effector CD8 T-cell response specific to at least two class I-restricted epitopes. The helper response was still observed when immunization was performed using LP6 plus a mixture of soluble capsid components released from detergent-disrupted virions. When mice were immunized with LP6 and each individual capsid component, i.e., hexon, penton base, or fiber, the results obtained suggested that hexon protein was responsible for the adjuvant effect exerted by disrupted Ad particles on the helper response to the immunogen (103).

Recombinant adenovirus (rAd) infection is one of the most effective and frequently employed methods to transduce dendritic cells (DC). Contradictory results have been reported concerning the influence of rAd on the differentiation and activation of DC. In one report it was shown, as a result of rAd infection, mouse bone marrow-derived immature DC upregulate expression of major histocompatibility complex class I and II antigens, costimulatory molecules (CD40, CD80, and CD86), and the adhesion molecule CD54 (ICAM-1). rAd-transduced DC exhibited increased allostimulatory capacity and levels of interleukin-6 (IL-6), IL-12p40, IL-15, gamma interferon, and tumor necrosis factor alpha mRNAs. These effects were not related to specific transgenic sequences or to rAd genome transcription. The rAd effect correlated with a rapid increase (1 h) in the NF-kappaB-DNA binding activity detected by electrophoretic mobility shift assays. rAd-induced DC maturation was blocked by the proteasome inhibitor Nalpha-p-tosyl-L-lysine chloromethyl ketone (TLCK) or by infection with rAd-IkappaB, an rAd-encoding the dominant-negative form of IkappaB. In vivo studies showed that after intravenous administration, rAds were rapidly entrapped in the spleen by marginal zone DC that mobilized to T-cell areas, a phenomenon suggesting that rAd also induced DC differentiation *in vivo*. These findings add a further pathway for the immunogenicity of rAd (104).

These data altogether indicate that even E1-E3 deleted recombinant adenoviral vectors, such as used in the ChAdOx1 nCov-19 (AZD1222, AstraZeneca) and Ad26.COV2.S (Johnson & Johnson/Janssen) adenoviral vector-based vaccines are powerful activators of the immune system, stimulating in particular the innate immune system at various stages

## Additive and synergistic effects of S protein and adenoviral vector

### Antigen mimickry

A recent publication has presented data which indicate that the severe side effects observed in rare cases may have to be attributed to adenoviral vaccines. Transcription of wildtype and codon-optimized Spike open reading frames enables alternative splice events that lead to C-terminal truncated, soluble Spike protein variants. These soluble Spike variants may initiate severe side effects when binding to ACE2-expressing endothelial cells in blood vessels. In analogy to the thromboembolic events caused by Spike protein encoded by the SARS-CoV-2 virus, the underlying disease mechanism has been termed the "Vaccine-Induced Covid-19 Mimicry" syndrome (VIC19M syndrome) vector-based vaccines (105).

## NF-κB promotes leaky expression of adenovirus genes in a replication incompetent adenovirus vector

The replication-incompetent adenovirus (Ad) vector is one of the most promising vectors for gene therapy, however, systemic administration of Ad vectors results in severe hepatotoxicities, partly due to the leaky expression of Ad genes in the liver. In one study it has been shown that NF-κB mediates the leaky expression of Ad genes from the Ad vector genome, and that the inhibition of NF-κB leads to the suppression of Ad gene expression and hepatotoxicities following transduction with Ad vectors. Activation of NF-κB by recombinant TNFα significantly enhanced the leaky expression of Ad genes. More than 50% suppression of the Ad gene expression was found by inhibitors of NF-κB signaling and siRNA-mediated knockdown of NF-κB. Similar results were found when cells were infected with wild-type Ad. Compared with a conventional Ad vector, an Ad vector expressing a dominant-negative IκBα (Adv-CADNIκBα), which is a negative regulator of NF-κB, mediated approximately 70% suppression of the leaky expression of Ad genes in the liver. Adv-CADNIκBα did not induce apparent hepatotoxicities. These results indicate that inhibition of NF-κB leads to suppression of Ad vector-mediated tissue damages *via* not only suppression of inflammatory responses but also reduction in the leaky expression of Ad genes (**106**).

Overall, these studies refered to in the present and previous chapters demonstrate that both, SARS-CoV-2 spike protein AND adenoviral vectors activate NF-κB pathway together with several additional signal transduction pathways.

## NF-κB activation is central to coagulation events

The effects of the NF-κB pathway activation on coagulopathies as described in a various studies described below may be relevant in VITT.

Plasminogen activator inhibitor-1 (PAI-1) is the major inhibitor of plasminogen activation and likely plays important roles in coronary thrombosis and arteriosclerosis. Tumor necrosis factor-alpha (TNFα) is one of many recognized physiological regulators of PAI-1 expression and may contribute to elevated plasma PAI-1 levels in sepsis and obesity. Although TNFα is a potent inducer of PAI-1 expression in vitro and in vivo, the precise location of the TNFα response site in the PAI-1 promoter was not determined. Transient transfection studies using luciferase reporter constructs containing PAI-1 promoter sequence up to 6.4 kb failed to detect a response to TNFα. Moreover, TNFα failed to induce expression of enhanced green fluorescent protein under the control of a 2.9-kb human PAI-1 promoter in transgenic mice. In contrast, a 5' distal TNFα-responsive enhancer of the PAI-1 gene located 15 kb upstream of the transcription start site containing a conserved NF-κB-binding site that mediates the response to TNFα. This newly recognized site was fully capable of binding NF-κB subunits p50 and p65, whereas overexpression of the NF-κB inhibitor IkappaB prevents TNFα-induced activation of this enhancer element (**107**).

In another study the development of procoagulant activity and monocyte activation in heparinized whole blood during extracorporeal circulation was studied. Anaphylatoxins, C3a and C5a appeared in the blood within 30 minutes of circulation. Circulated blood developed a marked potential for coagulation that reached maximal activity by 4 hours of circulation. This procoagulant activity was neutralized by anti-tissue factor antibody. Isolation of monocytes from circulated blood revealed that tissue factor expression is upregulated on the cell surface. Furthermore, they observed NF-κB nuclear translocation in monocytes from blood passing through the circuit, suggesting that tissue factor expression was due to monocyte stimulation and transcriptional activation of the tissue factor gene. Tissue factor expression resulted in an approximately 30-fold increase in thrombin generation. Monocyte NF-κB activation, monocyte tissue factor expression, thrombin generation, and the procoagulant activity of blood in extracorporeal circulation were all blocked by the proteasome inhibitor MG132 indicating that intravascular tissue factor expression during extracorporeal circulation of blood is due to NF-κB-mediated activation of monocytes (possibly by complement) (**108**).

In further study it was investigated whether complement components C3a and C5a regulate plasminogen activator inhibitor (PAI-1) in human macrophages. C5a increased PAI-1 up to 11-fold in human monocyte-derived macrophages (MDM) and up to 2.7-fold in human plaque macrophages. These results were confirmed at the mRNA level. Pertussis toxin or anti-C5aR/CD88 antibody completely abolished the effect of recombinant human C5a on PAI-1 production, suggesting a role of the C5a receptor. Experiments with anti TNFα antibodies and tiron showed that the effect of C5a was not mediated by TNFα or oxidative burst. Furthermore, C5a induced NF-κB binding to the cis element in human macrophages and the C5a-induced increase in PAI-1 was completely abolished by an NF-κB inhibitor indicating that C5a upregulates PAI-1 in macrophages via NF-κB activation (**109**).

Finally, platelets are megakaryocyte-derived fragments lacking nuclei and prepped to maintain primary hemostasis by initiating blood clots on injured vascular endothelia. Pathologically, platelets undergo the same physiological processes of activation, secretion, and aggregation yet with such pronouncedness that they orchestrate and make headway the progression of atherothrombotic diseases not only through clot formation but also via forcing a pro-inflammatory state. Indeed, NF-κB has been implicated in platelet survival and function (**110**)

**Process-related impurities in the ChAdOx1 nCov-19 vaccine**

ChAdOx1 nCov-19 vaccine lots have been analyzed by biochemical and proteomic methods. A recent preprint has shown that the vaccine, in addition to the adenovirus vector, contained substantial amounts of both human and non-structural viral proteins. The authors analyzed 3 different lots of the ChAdOx1 nCov-19 vaccine by SDS polyacrylamide gel electrophoresis (SDS-PAGE) followed by silver staining and compared the staining pattern of the separated proteins with those of HAdV-C5-EGFP, an adenovirus vector purified by CsCl ultracentrifugation followed by mass spectrometry. Based on intensity comparisons of LC/MS signals, they estimated that in one of three lots about 2/3 of the detected protein amounts were of human and 1/3 of virus origin, while the two other lots consisted of rather equal amounts of human and viral proteins. Beside the expected viral proteins that form the virion (hexon, penton base, IIIa, fiber, V, VI, VII,VIII, IX and others), also several non-structural viral proteins were detected at high abundancy although they are not part of the mature viral particle. Furthermore, peptides from more than 1000 different human proteins were detected being derived from the human vector production cell line. The detected proteins were derived from different cellular compartments including cytoplasm, nucleus, endoplasmic reticulum, Golgi apparatus and others. Intriguingly, from the human proteins found in the vaccine and beside several cytoskeletal proteins including Vimentin, Tubulin, Actin and Actinin, the group of heat shock proteins (HSPs) and chaperones stood out in abundancy. Among the top abundant proteins (including viral proteins) HSP 90-beta and HSP-90-alpha as cytosolic HSPs (with 9.5 % and 4.3 % of the total protein) and 3 chaperones of the endoplasmic reticulum (transitional endoplasmic reticulum ATPase, Endoplasmin and Calreticulin) were present. Extracellular HSPs are known to modulate innate and adaptive immune-responses, they can exacerbate pre-existing inflammatory condition, have been associated with autoimmunity and can even become target auf auto-immune responses themselves. They very efficiently initiate specific immune responses by receptor-mediated uptake of HSP-peptide complexes in antigen-presenting cells (APCs), mainly via CD91 and scavenger receptors. Furthermore, among the viral proteins detected, the adenoviral penton base is another candidate for inducing early toxicity via an RGD motive, present in a solvent-exposed loop of penton base, by interacting with integrins on cell membranes including of platelets (**111**).

**Discussion of an integrated mechanistic model for VITT following vaccination with adenovirus vector-based vaccines**

SARS-CoV-2 Spike protein is known to bind to the ACE2 receptor followed by endocytosis (**13**), whereas the primary mode of adenovirus (including replication defective E1/E3 deletion variants) is the high affinity binding of the fiber-knob to the CAR (coxsackie and adenovirus receptor), followed by interaction between the arginine-glycine-aspartate (RGD) motif within the viral penton base with the integrins on the cell surface, which facilitates then viral internalization (96). Beside the primary high affinity binding sites for cellular uptake both, SARS-CoV-2 and adenovirus have been described to bind and activate various Toll-like receptors (TLR) (76, 96, 112, 113). All these binding events of both, SARS-CoV-2 S-protein and adenoviral vector, have been described to activate – among various other – the NF-κB signaling pathway by activating IκB kinase (IKK) complexes, release of p50/p65 heterocomplex from the inhibitor complex with IκBa by proteasomal degradation of IκB, and translocation of the released and phosphorylated p50/p65 heterocomplex to the nucleus where the transcription of a huge panel of pro-inflammatory genes, including those for cytokines, such as TNFα, IL-1, IL-6, chemokines, such as MCP-1, MCP-3, adhesion molecule, such as ICAM-1 and VCAM-1, complement components and coagulation factors, such as PAI-1 (107) are induced. These processes are triggered in a broad variety of cell types, which express the respective receptors, including epithelial cells (e.g. in the lung), macrophages (including alveolar macrophages), and endothelial cells, found in all organs (76-79).

Expression of pro-inflammatory cytokines such as TNFα, IL-1, and IL-6, but also increased integrin expression (114, 115) can result in auto-amplifying positive feedback loops which affect also integrated systems such as complement and coagulation (39-44). The release of various chemokines, in particularly MCP-1 and IL-8 will stimulates migration and accumulation of various inflammatory and immune cells to the site. In particular neutrophils are attracted, activated by the pro-inflammatory cytokine/chemokine melieu and stimulated to interact with endothelial cells and platelets. Release of DNA chromatine nets, i.e. NETs have been described, which are considered one major mechanism for thromboses and occlusion of capillaries in COVID-19 (45-48).

While these events are known to accelerate and amplify in the case of massive viral infection by SARS-CoV-2 leading to COVID-19, the localized expression of the spike protein in case of vaccination is not expected to results in levels leading to a major systemic symptomatic picture. However, in the case of adenoviral vector vaccines, here comes the additive, and possibly in some cases, synergistic effects of the adenoviral vector itself. Beside additional activation of the NF-κB pathway – which may be considered a vaccine adjuvant effect – adenoviruses have the ability to bind and activate platelets and endothelial cells (97,98).

Adenoviral vectors can bind to circulating platelets which causes their activation and aggregation and subsequent entrapment in liver sinusoids. Moreover, activation of platelets is known to result in increased release of the PF4 – a tetramer with high positive surface charge, which on the one hand side can bind to the negatively charged glycoseaminoglycane structures on endothelial cells. As adenoviruses are known to activate platelets, it is plausible that the replication-deficient adenoviral vector could be directly responsible for the release of platelet-derived PF4. However, this hypothesis implies that significant amounts of vaccine particles would reach the bloodstream after intramuscular injection, which seems less likely. An alternative scenario would involve endothelial cells. Indeed, endothelial cells are efficiently transduced upon intramuscular injection. Transduced endothelial cells might be directly damaged by the spike protein that they synthesize. Platelets might then be recruited and activated by the spike protein expressed by endothelial cells. PF4 released by activated platelets may combine with anionic proteoglycans located on the surface of endothelial cells or are shed from endothelial cells. In such a scenario, both the adenoviral vector and the spike protein would contribute to the formation of immunogenic PF4 following vaccination with adenoviral vector-based COVID-19 vaccines.

One important central points of the VITT are the anti-PF4 antibodies found in the majority of patients with venous thromboses following SARS-CoV-2 adenoviral vector-based vaccines. The VITT induced antibodies are most probably induced by electrostatic binding of the positively charges PF4 tetramers to the negative surface of the adenoviral vector hexones. In this complex conformational changes occur – similar as reported for heparin induced HIT – which makes them highly immunogenic triggering the formation of anti-PF4 directed antibodies. Binding of these induced antibodies to the PF4 attached adenoviral surface will facilitate uptake of these complexes via FcgammaR mediated process (**90**). While macropages, monocytes, NK cells and DC will uptake antibody-PF4 (adenovirus) complexes via their different FCgamma receptors leading to activation of the cells, FcgammaR-induced complement activation, and degradation and sequestion of adenoviral vectors in reticulendothelial cells, also platelets are able to bind such complexes via their FcgammRIIA receptor leading to activation, release of PF4, aggregation of platelets, and resulting thromboses (**116**).

There is an additional mechanism that might accelerate the induction of anti-PF4 antibodies that is the supposed superantigen features of the Spike protein. Different from antigen specific activation of single T-cell clones in the case of conventional antigens, with the antigen presented in the MHC-TCR groove, superantigens have been decribed to bind outside this groove to MHC and /or TCR leading to the polyclonal activation (**63-65**). This, together with the anyway highly immunogenic PF4-adenovirus complexes, may facilitate the induction of PF4 specific antibodies. Notably, both, normal cell activation as well as superantigen-induced polyclonal T cell activation are dependent of NFκB pathway signaling (**71-75**).

Although originally being locally applied, a limited local dissemination of adenoviral vaccines e.g. via binding to endothelial cells (**117-119**) together with the described NF-κB-triggered leaky expression of adenovius genes in originally replication incompetent adenoviral vectors (**120**) may lead in rare cases to self-amplifying cascades leading to activated or damaged endothelial cells, activated and aggregated platelets, and activation of the coagulation systeme also at sites distant to the application site, i.e. systemic prothrombotic procoagulation events together with a corresponding (consumption) thrombocytopenia as observed in rare cases of adenovirus vector based SARS-CoV-2 vaccines

(121-126).

Additional activation of the described molecular and cellular pathways may occur by impurities such as the significant amounts of human heat shock proteins found in several ADT1222 vaccine preparations (111).

Together, these molecular and cellular mechanisms described in the previous sections may explain the higher prevalence of rare thromboses and thrombocytopenia following adenoviral vector-based anti SARS-CoV-2 vaccines compared to mRNA based anti SARS-CoV-2 vaccines.

## DATA AVAILABILITY STATEMENT

The data supporting the conclusions of this article will be made available by the authors, without undue reservation.

## AUTHOR CONTRIBUTIONS

RK contributed project idea, discussion of data, and writing of manuscript, literature search, and review of manuscript.

## DECLARATION OF CONFLICT OF INTEREST

No conflict of interest to be reported.

58. Arnold DM, Patriquin CJ, Nazy I. Thrombotic microangiopathies: a general approach to diagnosis and management. CMAJ. 2017;189(4):E153-E159.

59. Menter T, Haslbauer JD, Nienhold R, et al. . Postmortem examination of COVID-19 patients reveals diffuse alveolar damage with severe capillary congestion and variegated findings in lungs and other organs suggesting vascular dysfunction. Histopathology. 2020;77(2):198-209.

60. Nadkarni GN, Lala A, Bagiella E, Chang HL, Moreno PR, Pujadas E, et al. Anticoagulation, Bleeding, Mortality, and Pathology in Hospitalized Patients With COVID-19. J Am Coll Cardiol. 2020 Oct 20;76(16):1815-1826. doi: 10.1016/j.jacc.2020.08.041.

61. Logunov DY, Dolzhikova IV, Shcheblyakov DV, Tukhvatulin AI, Zubkova OV, Dzharullaeva AS, et al. Gam-COVID-Vac Vaccine Trial Group. Safety and efficacy of an rAd26 and rAd5 vector-based heterologous prime-boost COVID-19 vaccine: an interim analysis of a randomised controlled phase 3 trial in Russia. Lancet. 2021 Feb 20;397(10275):671-681. doi: 10.1016/S0140-6736(21)00234-8.

62. Hunter PR. Thrombosis after covid-19 vaccination. BMJ. 2021 Apr 14;373:n958. doi: 10.1136/bmj.n958..

63. Noval Rivas M, Porritt RA, Cheng MH, Bahar I, Arditi M. COVID-19-associated multisystem inflammatory syndrome in children (MIS-C): A novel disease that mimics toxic shock syndrome-the superantigen hypothesis. J Allergy Clin Immunol. 2021 Jan;147(1):57-59. doi: 10.1016/j.jaci.2020.10.008.

64. Cheng MH, Zhang S, Porritt RA, et al. Superantigenic character of an insert unique to SARS-CoV-2 spike supported by skewed TCR repertoire in patients with hyperinflammation. Proc Natl Acad Sci U S A. 2020;117(41):25254-25262. doi:10.1073/pnas.2010722117

65. Kouo T, Chaisawangwong W. SARS-CoV-2 as a superantigen in multisystem inflammatory syndrome in children. J Clin Invest. 2021 May 17;131(10):e149327. doi: 10.1172/JCI149327.

66. Scaglioni V, Soriano ER. Are superantigens the cause of cytokine storm and viral sepsis in severe COVID-19? Observations and hypothesis. Scand J Immunol. 2020 Dec;92(6):e12944.
34

page 36

**Figure Legends**

**Fig. 1**

SARS-CoV-2 Spike protein (expressed and released by tranfected cells following vaccination with either mRNA or adenoviral vector coding for Spike protein) binds to the ACE2 receptor followed by endocytosis. In contrast, high affinity binding of the adenovirus (including replication defective E1/E3 deletion variants) occurs via binding of the fiber-knob to the CAR (coxsackie and adenovirus receptor), followed by interaction between the arginine-glycine-aspartate (RGD) motif within the viral penton base with the integrins on the cell surface, which facilitates then viral internalization. Beside the primary high affinity binding sites for cellular uptake both, SARS-CoV-2 and adenovirus may bind and activate various Toll-like receptors (TLR) leading to activation – among various others – of the NFκB signaling pathway including activation of IKK complexes, release of p50/p65 complexs from the inhibitor complex with IκBa by proteasomal degradation of IκB, and translocation of the released and phosphorylated p50/p65 heterocomplex to the nucleus where the transcription of a huge variety of pro-inflammatory genes, including those for cytokines, such as TNFα, IL-1, IL-6, chemokines, such as MCP-1, MCP-3, adhesion molecule, such as ICAM-1 and VCAM-1, complement components and coagulation factors, such as PAI-1 are induced. These processes are triggered in a broad variety of cell types, which express the respective receptors, including epithelial cells (e.g. in the lung), macrophages (including alveolar macrophages), and endothelial cells, found in all organs.

Expression of pro-inflammatory cytokines such as TNFα, IL-1, and IL-6, and increased integrin expression can result in autoamplifying positive feedback loops, which may extend to other intergrated systems such as complement and coagulation. The release of chemokines, such as MCP-1 and IL-8 will attract migration and accumulation of inflammatory and immune cells to the site. In particular neutrophils are attracted, and activated by the pro-inflammatory cytokine/chemokine melieu and stimulated to interact with endothelial cells and platelets. Release of DNA chromatine nets, *i.e.* NETs are considered one major mechanism for thromboses and occlusion of capillaries.

While these events are known to accelerate and amplify in the case of massive viral infection by SARS-CoV-2 resulting in COVID-19, the localized expression of the spike protein in case of vaccination is not expected to results in levels inducing a major systemic symptomatic picture. However, in the case of adenoviral vector vaccines, additive and synergistic effects are possible. Beside an additional activation of the NFκB pathway adenoviruses have the ability directly to bind and activate platelets and endothelial cells. Adenovirus rapidly bind to circulating platelets, which causes their activation/ aggregation. Activation of platelets results in increased release of the PF4 – a tetramer with high positive surface charge, which on the one hand side can bind to the negartively charged glycoseaminoglycane on endothelial cells. Additionally, endothelial cells can be transduced upon intramuscular injection to express spike protein leading to further activation and damage of the endothelial cells.

One hallmark of the VITT are anti-PF4 antibodies detected in the majority of patients with venous thromboses and thrombocytopenia following SARS-CoV-2 adenoviral vector based vaccines. The VITT induced antibodies are most probably induced by electrostatic binding of the positively charges PF4 tetramers to the negative surface of the adenoviral vector hexones. In this complex conformational changes occur – similar as reported for heparin induced HIT – which makes them highly immunogenic resulting in formation of anti-PF4 directed antibodies. Binding of these induced antibodies to the PF4 attached adenoviral surface will in turn facilitate uptake of these complexes via FcgammaR mediated process. Macropages, monocytes, NK cells, and DC take up antibody-PF4 (adenovirus) complexes via different types of FCgamma-receptors leading to cellular activation, co-induced complement activation, and sequestion and degradation of adenoviral vectors in reticulendothelial cells. Also platelets are able to bind such complexes via their FcgammR2a receptor leading to activation, release of PF4, aggregation of platelets and resulting thromboses.

The supposed superantigen features of the Spike protein may present an additional mechanism to accelerate the induction of anti-PF4 antibodies. Different from antigen specific activation of single T-cell clones in the case of conventional antigens, with the antigen presented in the MHC-TCR groove, superantigens bind outside this groove to MHC and /or TCR leading to the polyclonal activation. This, together with the highly immunogenic PF4-adenovirus complexes, may further facilitate the induction of PF4 specific antibodies. Notably, both, normal cell activation as well as superantigen-induced polyclonal T cell activation are dependent of NFκB pathway signaling.

Although originally being locally applied, a starting local dissemination of adenoviral vaccines via binding to endothelial cells and platelets together with the described NFκB-promoted leaky expression of adenovius genes in originally replication-incompetent adenoviral vectors can lead in rare cases to self-amplifying cascades leading to activated damaged endothelial cells, activated and aggregated platelets, and activation of the coagulation systeme even at sites distant to the application site, resulting in systemic pro-thrombotic procoagulation events going along with corresponding (consumption) thrombocytopenia as observed in rare cases of adenovirus vector based SARS-CoV-2 vaccines.

**Fig. 1**

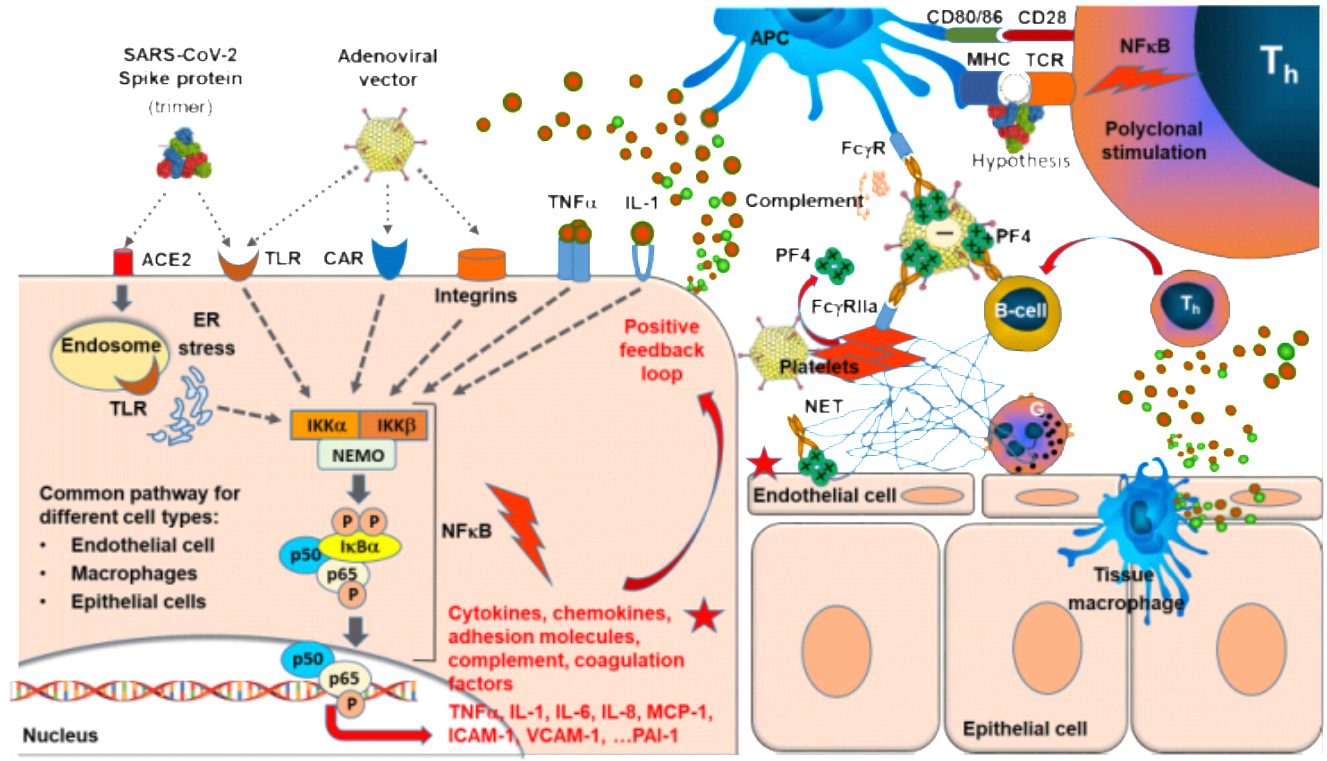